# Long-term data storage in diamond

Siddharth Dhomkar[1,†], Jacob Henshaw[1,2,†], Harishankar Jayakumar[1], Carlos A. Meriles[1,2,*]

[1]*Dept. of Physics, CUNY-City College of New York, New York, NY 10031, USA.*
[2]*CUNY-Graduate Center, New York, NY 10016, USA.*
[†] *Equally contributing authors.*

The negatively-charged nitrogen-vacancy (NV⁻) center in diamond is the focus of widespread attention for applications ranging from quantum information processing to nanoscale metrology. Although most work so far has focused on the NV⁻ optical and spin properties, control of the charge state promises complementary opportunities. One intriguing possibility is the long-term storage of information, a notion we hereby introduce using NV rich, type-1b diamond. As a proof of principle, we use multi-color optical microscopy to read, write, and reset arbitrary data sets with 2-D binary bit density comparable to present digital-video-disk (DVD) technology. Leveraging on the singular dynamics of NV⁻ ionization, we encode information on different planes of the diamond crystal with no cross talk, hence extending the storage capacity to three dimensions. Further, we correlate the center's charge state and nuclear spin polarization of the nitrogen host, and show that the latter is robust to a cycle of NV⁻ ionization and recharge. In combination with super-resolution microscopy techniques, these observations provide a route towards sub-diffraction NV charge control, a regime where the storage capacity could exceed present technologies.

## INTRODUCTION

Diamond is a unique platform material whose extreme properties and multi-functionalities are enabling an ever growing set of applications ranging from the fabrication of long-lasting machining and cutting tools, to biomedical and low-wear coatings, to efficient heat sinks for high-power electronics. Diamond typically contains impurities and other defects whose varying concentration and composition give gems their signature colors. An example of emerging importance is the negatively charged nitrogen-vacancy center (NV⁻), a spin-1 complex formed by a substitutional nitrogen atom adjacent to a vacant site. These paramagnetic centers can be located individually using confocal microscopy, initialized via optical pumping, and read out through spin dependent photoluminescence measurements[1]. Optical access coupled to single electron spin control and millisecond-long coherence spin lifetimes under ambient conditions[2] has led to recent demonstrations of entanglement and basic quantum logic[3-7] as well as various forms of nanoscale sensing[8-11].

Here we use multi-color optical microscopy to locally convert the charge state of NVs within a dense ensemble from negative to neutral, and correspondingly alter the NV fluorescence emission from bright to dark. This change is reversible, long-lasting, and robust to weak illumination, thus serving as an alternate platform for three-dimensional (3D) information storage. To demonstrate this notion, we write, read, erase, and rewrite data, which here take the form of stacked two-dimensional images. Though presently limited by light diffraction, access to the NV electron and nuclear spin degrees of freedom could be potentially exploited to reduce the volume per bit. As a first step in this direction, we use a charge-to-spin conversion protocol to polarize the nuclear spin of the nitrogen host conditional on the NV charge state, and subsequently show that this polarization remains virtually unchanged throughout a cycle of NV$^-$ ionization and recharge.

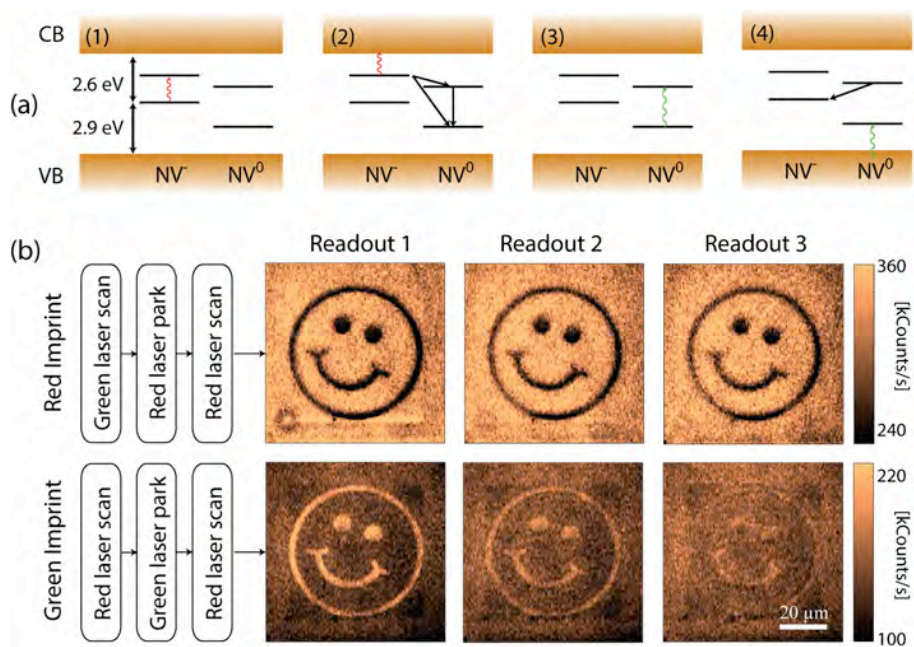

**Fig. 1. Charge manipulation and readout in diamond.** (a) Energy diagram for NV⁻ and NV⁰. In (1) and (2) the successive absorption of two photons (wavy arrows) of energy equal or greater than 1.95 eV (637 nm) propels the excess electron of an NV⁻ into the conduction band leaving the defect in the neutral ground state (solid arrows). In (3) and (4), an NV⁰ consecutively absorbs two photons of energy equal or greater than 2.16 eV (575 nm) transforming into NV⁻. CB and VB are the conduction and valence bands, respectively. (b) (Upper row) A binary pattern on an NV⁻-rich background is imprinted via spatially selective red illumination (632 nm, 350 μW, 100 ms per pixel). (Lower row) Starting from an NV⁻-depleted background the pattern results from selective illumination with green laser light (532 nm, 30 μW, 5 ms per pixel). From left to right, images are the result of three successive readouts of the imprint via a red scan (200 μW and 150 μW for the upper and lower rows, respectively). In all cases, the image size is 100x100 pixels, and the integration time is 1 ms per pixel.

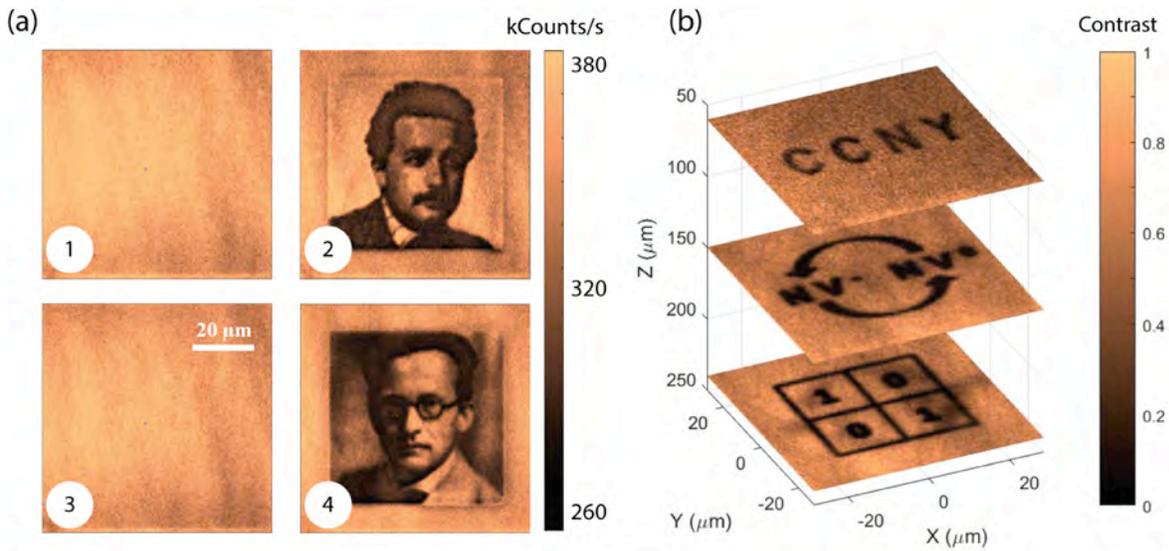

**Fig. 2. Diamond as a three-dimensional read/write memory.** (a) Starting from a blank ensemble of NV$^-$ centers (1), information can be written (2), erased (3), and rewritten (4). In (1) and (3) a green laser scan (1 mW at 1 ms per pixel) is used to reset the target plane to a bright state. In (2) and (4) images are imprinted via a red laser scan with a variable exposure time per pixel (from 0 to 50 ms). Note the gray scale in the resulting images corresponding to multi-valued (as opposed to binary) encoding. The same scale bar applies to all four images. (b) Information can be stored and accessed in three dimensions, as demonstrated for the case of a three-level stack. Observations over a period of a week show no noticeable change in these patterns for a sample kept in the dark. In (a) and (b) readout is carried out via a red laser scan (200 µW at 1 ms per pixel). The image size is 150x150 pixels in all cases.

## RESULTS

### Storing and retrieving arbitrary data sets in three dimensions

The physical mechanisms underlying NV charge dynamics are presented in Fig. 1a: Green illumination (e.g., at 532 nm) ionizes the NV$^-$ via the consecutive absorption of two photons, thus transforming the NV$^-$ into NV$^0$ (i.e., a neutrally charged NV). Conversely, green light can drive a neutral NV into its excited state where absorption of an electron from the valence band reconverts NV$^0$ back into NV$^-$ (Fig. 1a). Therefore, an NV center exposed to green light dynamically alters its charge state at a rate that depends on the illumination intensity[12,13]. This behavior changes with the use of red light (e.g., 632 nm) because photons of this wavelength can only excite NV$^-$ but not NV$^0$. Consequently, strong red illumination ionizes NV$^-$ to produce NV$^0$ but the back-conversion process is largely inhibited.

In our experiments we use a type-1b diamond crystal with an approximate NV concentration of 0.4 ppm. Two intuitive forms of charge patterning are presented in Fig. 1b: Upon initializing the focal plane into NV$^-$ (upper row) we convert into NV$^0$ select portions by successively parking a strong red beam at the desired pixels for a pre-defined time interval. Given the near quadratic dependence of the ionization rate on the illumination intensity[13], the resulting NV charge map can be revealed via a weak red laser scan. In this regime charge ionization during readout is minimal and the fluorescence — brighter in NV$^-$-rich areas — correlates with the NV$^-$ concentration. The lower row in Fig. 1b illustrates the converse approach where patterning is attained by parking a green beam on a 'bleached' (i.e., NV$^-$-deprived) plane. Exposure to green light locally reconverts NV$^0$ into NV$^-$ and subsequent fluorescence imaging — via a weak red scan — unveils the expected bright pattern on an otherwise dark background.

Both encoding protocols yield comparable pixel definition (about 0.8 µm, here defined by the numerical aperture $NA$=0.42 of the objective, see Fig. S1). Green or red imprints, however, respond differently to multiple red laser readouts (middle and left columns in Fig. 1b). Both exhibit a gradual loss of contrast but the impact is substantially stronger on the green imprint. Remarkably, observations on test patterns over a period of a week show no noticeable change provided the diamond crystal is kept in the dark. Thus, data storage in diamond must be viewed as semi-permanent in the sense that a 'refresh' protocol is required, conditional on the number of readouts but independent on the total elapsed time.

To derive a more quantitative metric we compare the fluorescence response from an arbitrary (but fixed) site of the diamond crystal to multiple readouts (Fig. S2). We find that red imprinting not only features a slower fluorescence decay but also that the relative contrast between 'bright' and 'dark' remains high over tens of readouts. This is not the case for a green imprint where the contrast first vanishes and then inverts. The physics at play is complex and more investigation will be needed to gain a fuller understanding. Initial work, however, indicates that the local N$^+$ content — higher when green light is present during the encoding process — plays an important role[14].

Unlike photo-refractive polymers[15] — prone to degradation upon repeated light exposure — or gold nanorods[16,17] — undergoing a permanent photo-induced shape change — the charge state of the NV center can be reversibly altered with no accumulated effect, hence allowing one to erase and rewrite information a virtually limitless number of times. A proof-of-principle demonstration is presented in Fig. 2a (see also Fig. S3): After resetting the NV$^-$ content via a strong green laser scan (Step 1), we proceed to write the focal plane through a red imprint, which we then expose via a weak red scan (Step 2); the same protocol is then repeated to encode and readout a new, different pattern (Steps 3 and 4). Note that unlike Fig. 1 — where the brightness in each pixel takes one of two possible values — the images of Fig. 2a are imprinted using a variable exposure time per pixel. In the present case we bin the illumination times into five different durations, which correspondingly lead to discernible levels of fluorescence, i.e., the equivalent of a multi-valued bit (Fig. S1). The result is a concomitant boost of the information density, here illustrated via the gray-scale images. The number of levels is largely defined by the signal-to-noise ratio (SNR) of the optical detection, which, in turn, grows with the square root of the readout time and NV density. In practice, considerations such as background noise and sample homogeneity must also be taken into account. For the conditions herein, up to 8 different levels seem realistic (see Fig. S1) though more are conceivable, e.g., if the sample is engineered to host a higher NV concentration.

Since the illumination intensity decays with the inverse square of the distance to the focal plane, it is possible to imprint the diamond selectively at a given depth without altering the information stored elsewhere. A demonstration is presented in Fig. 2b, where we write NV charge maps on three stacked planes approximately 90 μm apart from each other. Given the thickness of the diamond sample we used (200 μm), these results indicate minimal optical aberrations throughout the crystal. On the other hand, the separation between planes — largely defined by the beam shape near the focal plane — could be reduced by resorting to beam shaping techniques. In particular, a spatial light modulator could be used to adjust the optical wave front to reduce axial elongation in the beam profile[18].

Ultimately, the inter-plane separation results from a tradeoff between various parameters including the required level of contrast, writing speed, and light intensity. For example, better in-plane localization is attained in the limit of low laser power, where the NV ionization rate responds quadratically to the illumination intensity[13], but the encoding time per pixel is comparatively longer. Faster writing speed can be reached with stronger laser power, but saturation of the first excited state gradually makes the NV ionization rate transition from quadratically- to linearly-dependent on the intensity, with the corresponding reduction of the in-plane localization[12]. For a given laser power, a similar consideration applies to the light exposure time and fluorescence contrast, the latter improving with longer imprint times at the expense of a larger pixel volume (see also Fig. S1). Note, however, that this tradeoff has a lesser impact on data density if the brighter fluorescence of larger pixels is binned into discrete levels to produce multi-valued bits, as discussed above (Fig. 2a).

The absolute write and read times per pixel — either comparable to or greater than 1 ms — presently make NV storage comparatively slow for practical applications though there seems to be considerable room for improvement. The most obvious route to faster writing makes use of stronger illumination intensities, though at the expense of higher power consumption and system complexity. For a constant average laser power, pulsed excitation may prove beneficial given the quadratic response of NV ionization upon green or red illumination. Along the same lines, different excitation colors can exhibit markedly different ionization efficiencies (see, e.g., conditions in Fig. 1 for red and green imprinting) thus calling for a systematic characterization as a function of the excitation wavelength. In particular, we show below that $NV^-$ can be efficiently ionized by blue illumination (directly exciting the excess electron into the conduction band) though further work will be needed for a vis-à-vis comparison between one- and two-photon ionization in type 1b diamond. Although some of the same considerations also impact readout speed, the latter is mainly defined by SNR limitations, which, perhaps, could be ameliorated by increasing the NV content.

**Towards super-resolution data storage**

While the spatial resolution of a diamond memory — or, for that matter, any other optical memory[15-19] — is inherently influenced by light diffraction, a question of interest is whether the latter sets a fundamental limit for manipulating the NV charge. Super-resolution methods have already been applied to image NV centers with spatial accuracy of up to ~6 nm[20], approximately one hundredth of the excitation wavelength. However, storing and accessing information with sub-diffraction discrimination would require that the NV charge state be preserved during the write and readout processes, a condition difficult to meet with existing super-resolution imaging methods. This incompatibility is apparent in schemes such as Stochastic Optical Reconstruction Microscopy[21] (STORM) or Photo-Activated Localization Microscopy (PALM)[22], where spatial resolution is attained by randomly activating a small fraction of fluorophores while most of the ensemble remains in the dark state[23].

Sub-diffraction imaging strategies that deterministically drive NVs into a non-fluorescing state are not exempt from problems. For example, in Charge State Depletion (CSD) microscopy[24], fluorescence is recorded via the use of weak, non-ionizing illumination following the successive application of a green, Gaussian beam and a concentric, doughnut-shaped, red beam. The former brings most NVs within the focal area into the negative, bright state whereas the latter selectively transforms peripheral NVs into the dark, neutrally charged state. CSD microscopy, therefore, is unsuited for high-density, sub-diffraction recording because any given 'write' operation initializes the charge state of NVs proximal to the target. This same drawback also applies to Stimulated Emission Depletion[20] (STED) microscopy and related techniques[25,26] because uncontrolled NV ionization is at least as likely as stimulated emission during application of the strong STED beam.

The ability to manipulate the NV spin degrees of freedom provides a versatile route to circumvent these problems. For example, since nuclear spins are relatively well isolated, data loss during a super-resolution read/write could be eluded via the use of a charge-to-spin (CTS) conversion scheme, where the nuclear spins of all NVs within the laser focal spot are polarized conditional on the initial NV charge state. This route exploits the NV nuclear spins as ancillary memories to temporarily store the initial charge state of all illuminated NVs during a laser read/write. Using a doughnut beam to separately address the group of NVs surrounding the target, the original charge state can be subsequently re-established via spin-to-charge (STC) conversion[27]. Here we are assuming that the experimental conditions are chosen so that the full protocol — including CTS, target read/write, and STC — takes place on a time scale shorter than the spin lattice relaxation time of the NV nuclear host.

An initial proof of concept containing key ingredients necessary for the realization of the above approach is presented in Fig. 3: Upon preparing the NVs into the negatively charged state, we use an optical pumping scheme[28] to initialize the $^{14}N$ spin of the host nitrogen atom — a system of spin number $I=1$ — into $m_I=0$ (where $m_I$ denotes the nuclear spin quantum projection along a direction coincident with the NV axis). This scheme uses a train of laser, microwave (mw), and radio-frequency (rf) pulses to drive the $^{14}$N spin into a desired final state (Fig. 3a and Section S1 of the Supplementary Material). Optically detected magnetic resonance (ODMR) of the NV$^-$ electronic spin (right column in Fig. 3b) reveals a hyperfine-split spectrum with a prominent central peak surrounded by two weak satellites, from where we estimate the level of nuclear spin polarization at about 80%.

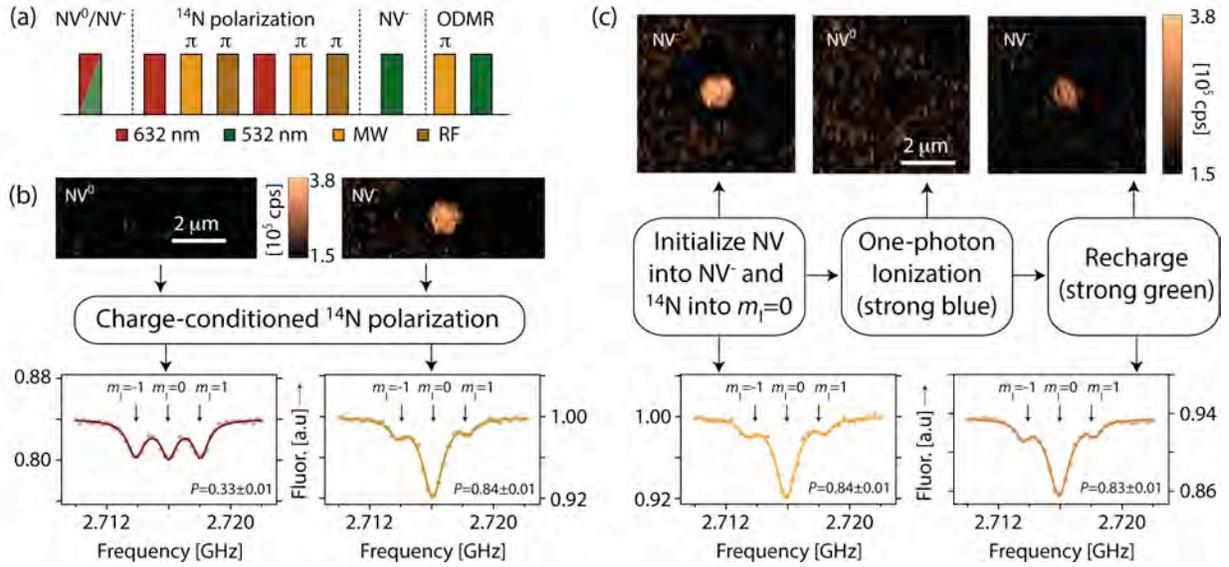

**Fig. 3. Combined NV charge and spin manipulation.** (a) Charge-conditional initialization of the $^{14}$N nuclear spin host into $m_I=0$ is attained via spin transfer from the optically polarized NV$^-$ electronic spin (see also Figs. S4 and S5); MW and RF denote microwave and radio-frequency pulses, respectively. Initialization into NV$^-$ (NV$^0$) is attained by applying (or not) a green laser pulse (532 nm, 1 mW, 10 μs) on an NV$^-$-depleted background (632 nm, 250 μW, 50 ms per pixel). Following $^{14}$N polarization, (unconditional) reconversion into NV$^-$ is attained via a green laser pulse (532 nm, 1 mW, 3 μs). The durations of the microwave and radio-frequency pulses are 440 ns and 28 μs, respectively. (b) Measured NV$^-$ ODMR spectra after application of the pulse sequence in (a). The upper images (632 nm, 250 μW, 1 ms per pixel) show the NV$^-$ fluorescence in a vicinity of the probed sample spot (coincident with the image center) after charge initialization. (c) With the $^{14}$N spin in the $m_I=0$ state, negatively charged NVs undergo a cycle of ionization and recharge. The pulse protocol is identical to that in (a) except that a blue laser pulse (450 nm, 400 μW, 30 μs) is introduced after nuclear spin polarization temporarily converting NV$^-$ into NV$^0$, as indicated by the upper fluorescence images. Comparison of the ODMR spectra before (left) and after (right) application of the ionization/recharge cycle shows nearly no change of the $^{14}$N spin polarization. In (b) and (c) every point in the ODMR spectra corresponds to $10^5$ consecutive averages; solid lines are Lorentzian fits to the three $^{14}$N hyperfine peaks; $P$ denotes the fractional area under the central peak.

Because both the electronic and spin energy levels depend on the center's charge state, the spin pumping protocol has no impact on the nuclear spins of neutral NVs, which consequently remain unpolarized. Such charge selectivity is confirmed by interrogating NV centers initially prepared in the neutral state, and subsequently converted to NV$^-$ prior to ODMR inspection (left column in Fig. 3b); as expected, the spectrum displays in this case three peaks of comparable amplitude, indicative of equal $^{14}$N spin populations in all three projections. Conditional $^{14}$N spin polarization thus amounts to 'charge-to-spin' conversion, with fidelity ultimately limited by the chosen nuclear spin initialization protocol.

To recreate the impact of super-resolution schemes on the charge state of NVs near the target, we impose a cycle of forced ionization and recharge (Fig. 3c and Section S2 of the Supplementary Material). Departing from the experiments of Figs. 1 and 2, NV$^-$ ionization is carried out this time with the aid of a femtosecond laser tuned to emit at 450 nm. Unlike red or green illumination — where the NV undergoes a two-step process, Fig. 1a — one-photon excitation in the blue directly propels the NV$^-$ excess electron into the conduction band, hence avoiding light-induced nuclear spin depolarization via level mixing in the first excited state[29]. To recharge the NV$^0$ we use a strong green laser pulse whose duration is optimized to yield one-directional charge conversion into the NV$^-$ state with minimum $^{14}$N spin depolarization. Comparison of the ODMR spectra before and after the ionization-and-recharge cycle (lower left and right plots in Fig. 3c, respectively) shows virtually no change in the $^{14}$N spin polarization, which demonstrates data protection against photo-induced charge conversion (see also Figs. S4 and S5).

## DISCUSSION

Further work will be necessary to gain charge control of individual or small groups of NVs with sub-diffraction resolution, starting with the implementation of more efficient charge-to-spin conversion protocols, here limited by the fidelity of the chosen nuclear spin polarization scheme. Though the present experiments are carried out at only ~5.5 mT, greater nuclear spin resilience to photo-ionization is possible at higher magnetic fields (>200 mT) where state-mixing-driven nuclear spin relaxation in the excited state is significantly reduced[30]. By the same token, greater nuclear spin polarization contrast between neutral and negative NVs is feasible if the charge-to-spin protocol is improved so as to polarize complementary nuclear spin projections depending on the original charge state (e.g., $m_I=-1$ for NV$^-$ and $m_I=+1$ for NV$^0$).

Similar considerations apply to the converse operation, namely, the 'spin-to-charge' transformation presently attainable at only moderate fidelities[27]. Given the high spin selectivity of the inter-system crossing at room temperature, more efficient spin-to-charge conversion is conceivable if the existing protocol is modified so as to induce ionization by one-photon excitation from the ground singlet state. Alternatively, it may be possible to exploit the longer electronic state lifetimes of NVs at low temperatures, though at the expense of a more complex experimental setup.

Circumventing the limitations inherent to storage media confined to two dimensions[31,32], the ideas discussed herein can be extended to include other defects also acting as electron traps. Such centers could be exploited, e.g., for error correction so as to mitigate charge instabilities from electron tunneling between neighboring NV centers, or between NV centers and surrounding substitutional nitrogen (likely in crystals with high NV and nitrogen concentration[33]). By the same token, material platforms other than NVs in diamond may also be exploited for high-density data storage; examples include the silicon-carbon di-vacancy[34] and the silicon vacancy[35] in SiC as well as select rare earth ions in garnets[36], all of which exhibit controllable charge and electronic/ nuclear spin degrees of freedom.

## MATERIALS AND METHODS

### Diamond crystal

The sample is a type-1b [111] diamond from Diamond Delaware Knives (DDK). Prior characterization via infrared spectroscopy[14] is consistent with the presence of substitutional nitrogen at a concentration of approximately 40 ppm; the estimated NV content is 0.4 ppm. Absorption near 1282 cm$^{-1}$ suggests that A-centres — formed by two adjacent nitrogens — are, if at all present, at trace concentrations. Optical spectroscopy confirms that the collected fluorescence originates almost exclusively from NV centres. A distinctive peak at ~737 nm reveals the presence of silicon-vacancy (SiV) centres; from the peak amplitude we estimate the SiV-NV ratio to be about 0.6 %.

### NV magnetic resonance and optical microscopy

For our experiments we use a custom-made, multi-color microscope. A 13 mW helium-neon laser and a 2 W cw solid-state laser serve as the sources of red (632 nm) and green (532 nm) light, respectively. Excitation in the blue (450 nm) is provided by a tunable ultrafast laser (Coherent Mira) and a frequency doubler generating 120-fs-long pulses at a repetition of 76 MHz; the average power at 450 nm is 400 μW. All laser beams are coupled into a 0.42 numerical aperture objective, which also collects the outgoing sample fluorescence. The illumination timing is set independently with the aid of acousto-optic modulators (AOM); a servo-controlled, two-mirror galvo system is used for sample scanning. Sample fluorescence in the range 650 – 850 nm is detected after a dichroic mirror and notch filters by a solid-state avalanche photo detector (APD).

Control of the NV$^-$ electronic and nuclear spin is carried out via the use of microwave (MW) and radio-frequency (RF) pulses produced by four signal generators (Rohde&Schwarz SMB100A, Rohde&Schwarz SMV03, Agilent E4433B, and Tektronix AFG3102). A 20-μm-diameter copper wire overlaid on the diamond surface serves as the simultaneous source of the MW and RF fields. Upon amplification, the typical duration of a MW (RF) inversion pulse is 500 ns (30 μs). All magnetic resonance experiments are carried out in the presence of an 5.5 mT magnetic field emanating from a permanent magnet in the sample vicinity. The magnetic field is oriented so as to coincide with the sample crystal normal, i.e., the [111] axis. A pulse generator (PulseBlaster ESR-PRO) controls the timing of all laser, MW, and RF pulses. All experiments are carried out under ambient conditions.

## SUPPLEMENTARY MATERIALS

Section S1. Charge-conditioned polarization of the $^{14}$N spin.
Section S2. Impact of NV$^-$ ionization and recharge on $^{14}$N spin polarization.
Fig. S1. Spatial resolution of NV charge patterning.
Fig. S2. Impact of multiple readouts on NV fluorescence contrast.
Fig. S3. NV response upon multiple read/write cycles.
Fig. S4. Impact of NV$^-$ ionization on the $^{14}$N nuclear spin polarization.
Fig. S5. Impact of NV$^-$ recharge on the $^{14}$N nuclear spin polarization.

## REFERENCES AND NOTES

[1] F. Jelezko, T. Gaebel, I. Popa, A. Gruber, J. Wrachtrup, Observation of Coherent Oscillations in a Single Electron Spin, *Phys. Rev. Lett.* **92**, 076401 (2004).
[2] G. Balasubramanian, P. Neumann, D. Twitchen, M. Markham, R. Kolesov, N. Mizuochi, J. Isoya, J. Achard, J. Beck, J. Tissler, V. Jacques, P. Hemmer, F. Jelezko, J. Wrachtrup, Ultralong spin coherence time in isotopically engineered diamond. *Nature Mater.* **8**, 383-387 (2009).
[3] M.V. Gurudev Dutt, L. Childress, L. Jiang, E. Togan, J. Maze, F. Jelezko, A.S. Zibrov, P.R. Hemmer, M.D. Lukin, Quantum Register Based on Individual Electronic and Nuclear Spin Qubits in Diamond. *Science* **316**, 1312-1316 (2007).
[4] P. Neumann, N. Mizuochi, F. Rempp, P. Hemmer, H. Watanabe, S. Yamasaki, V. Jacques, T. Gaebel, F. Jelezko, J. Wrachtrup, Multipartite Entanglement Among Single Spins in Diamond. *Science* **320**, 1326-1329 (2008).
[5] P. Neumann, R. Kolesov, B. Naydenov, J. Beck, F. Rempp, M. Steiner, V. Jacques, G. Balasubramanian, M.L. Markham, D.J. Twitchen, S. Pezzagna, J. Meijer, J. Twamley, F. Jelezko, J. Wrachtrup, Quantum register based on coupled electron spins in a room-temperature solid. *Nat. Phys.* **6**, 249-253 (2010).
[6] P.C. Maurer, G. Kucsko, C. Latta, L. Jiang, N.Y. Yao, S.D. Bennett, F. Pastawski, D. Hunger, N. Chisholm, M. Markham, D.J. Twitchen, J.I. Cirac, M.D. Lukin, Room-temperature quantum bit memory exceeding one second. *Science* **336**, 1283-1286 (2012).
[7] E. Togan, Y. Chu, A.S. Trifonov, L. Jiang, J. Maze, L. Childress, M.V.G. Dutt, A.S. Sørensen, P.R. Hemmer, A.S. Zibrov, M.D. Lukin, Quantum entanglement between an optical photon and a solid-state spin qubit. *Nature* **466**, 730-734 (2010).
[8] T. Staudacher, F. Shi, S. Pezzagna, J. Meijer, J. Du, C.A. Meriles, F. Reinhard, J. Wrachtrup, Nuclear magnetic resonance spectroscopy on a (5-nanometer)$^3$ volume of liquid and solid samples. *Science* **339**, 561-563 (2013).
[9] H.J. Mamin, M. Kim, M.H. Sherwood, C.T. Rettner, K. Ohno, D.D. Awschalom, D. Rugar, Nanoscale nuclear magnetic resonance with a nitrogen-vacancy spin sensor. *Science* **339,** 557-560 (2013).
[10] T. Häberle, D. Schmid-Lorch, F. Reinhard, J. Wrachtrup, Nanoscale nuclear magnetic imaging with chemical contrast. *Nat. Nanotech.* **10**, 125-128 (2015).
[11] T.M. Staudacher, N. Raatz, S. Pezzagna, J. Meijer, F. Reinhard, C.A. Meriles, J. Wrachtrup, Probing molecular dynamics at the nanoscale via an individual paramagnetic center. *Nat. Commun.* **6**, 8527 (2015).
[12] G. Waldherr, J. Beck, M. Steiner, P. Neumann, A. Gali, Th. Frauenheim, F. Jelezko, J. Wrachtrup, Dark States of Single Nitrogen-Vacancy Centers in Diamond Unraveled by Single Shot NMR. *Phys. Rev. Lett.* **106**, 157601 (2011).
[13] N. Aslam, G. Waldherr, P. Neumann, F. Jelezko, J. Wrachtrup, Photo-induced ionization dynamics of the nitrogen vacancy defect in diamond investigated by single-shot charge state detection. *New J. Phys.* **15**, 013064 (2013).
[14] H. Jayakumar, J. Henshaw, D. Pagliero, A. Laraoui, N.B. Manson, R. Albu, M.W. Doherty, C.A. Meriles, Optical patterning of trapped charge in nitrogen-doped diamond. *Nat. Commun.* **7**, 12660 (2016).
[15] D. Day, M. Gu, A. Smallridge, Use of two-photon excitation for erasable–rewritable three-dimensional bit optical data storage in a photo refractive polymer. *Optics Lett.* **24**, 948-950 (1999).
[16] P. Zijlstra, J.W.M. Chon, M. Gu, Five-dimensional optical recording mediated by surface plasmons in gold nano rods. *Nature* **459**, 410-413 (2009).
[17] Z. Gan, Y. Cao, R.A. Evans, M. Gu, Three-dimensional deep sub-diffraction optical beam lithography with 9 nm feature size. *Nat. Commun.* **4**, 2061 (2013).
[18] Y-C. Chen, P.S. Salter, S. Knauer, L. Weng, A.C. Frangeskou, C.J. Stephen, P.R. Dolan, S. Johnson, B.L. Green, G.W. Morley, M.E. Newton, J.G. Rarity,

M.J. Booth, J.M. Smith, Laser writing of coherent colour centres in diamond. arXiv:1606.05109 (2016).

[19] M. Gu, X. Li, Y. Cao, Optical storage arrays: a perspective for future big data storage. *Light Sci. Appl.* **3**, e177 (2014).

[20] E. Rittweger, K. Young Han, S.E. Irvine, C. Eggeling, S.W. Hell, STED microscopy reveals crystal colour centres with nanometric resolution. *Nat. Photonics* 3, 144-147 (2009).

[21] M.J Rust, M. Bates, X. Zhuang, Sub-diffraction-limit imaging by stochastic optical reconstruction microscopy (STORM). *Nat. Meth.* **3**, 793-795 (2006).

[22] E. Betzig, G.H. Patterson, R. Sougrat, O.W. Lindwasser, S. Olenych, J.S. Bonifacino, M.W. Davidson, J. Lippincott-Schwartz, H.F. Hess, Imaging intracellular fluorescent proteins at nanometer resolution. *Science* **313**, 1642-1645 (2006).

[23] M. Pfender, N. Aslam, G. Waldherr, P. Neumann, J. Wrachtrup, Single-spin stochastic optical reconstruction microscopy. *Proc. Natl. Acad. Sci. USA* **111**, 14669-14674 (2014).

[24] X. Chen, C. Zou, Z. Gong, C. Dong, G. Guo, F. Sun, Subdiffraction optical manipulation of the charge state of nitrogen vacancy center in diamond. *Light: Science & Applications* **4**, e230 (2015).

[25] E. Rittweger, D. Wildanger, S.W. Hell, Far-field fluorescence nanoscopy of diamond color centers by ground state depletion. *Eurphys. Lett.* **86**, 14001 (2009).

[26] K.Y. Han, S.K. Kim, C. Eggeling, S.W. Hell, Metastable Dark States Enable Ground State Depletion Microscopy of Nitrogen Vacancy Centers in Diamond with Diffraction-Unlimited Resolution. *Nano Lett.* **10**, 3199-3203 (2010).

[27] B.J. Shields, Q.P. Unterreithmeier, N.P. de Leon, H. Park, M.D. Lukin, Efficient Readout of a Single Spin State in Diamond via Spin-to-Charge Conversion. *Phys. Rev. Lett.* **114**, 136402 (2015).

[28] D. Pagliero, A. Laraoui, J. Henshaw, C.A. Meriles, Recursive polarization of nuclear spins in diamond at arbitrary magnetic fields. *Appl. Phys. Lett.* **105**, 242402 (2014).

[29] L. Jiang, M.V. Gurudev Dutt, E. Togan, L. Childress, P. Cappellaro, J.M. Taylor, M.D. Lukin, Coherence of an optically illuminated single nuclear spin qubit. *Phys. Rev. Lett.* **100**, 073001 (2008).

[30] P. Neumann, J. Beck, M. Steiner, F. Rempp, H. Fedder, P.R. Hemmer, J. Wrachtrup, F. Jelezko, Single-Shot Readout of a Single Nuclear Spin. *Science* **329**, 542-544 (2010).

[31] S. Loth, S. Baumann, C.P. Lutz, D.M. Eigler, A.J. Heinrich, Bistability in atomic-scale antiferromagnets. *Science* **335**, 196-199 (2012).

[32] C.R. Moon, L.S. Mattos, B.K. Foster, G. Zeltzer, H.C. Manoharan, Quantum holographic encoding in a two-dimensional electron gas. *Nat. Nanotech.* **4**, 167-172 (2009).

[33] N. Manson and J. Harrison, Photo-ionization of the nitrogen-vacancy center in diamond. *Diamond Relat. Mater.* **14**, 1705-1710 (2005).

[34] W.F. Koehl, B.B. Buckley, F.J. Heremans, G. Calusine, D.D. Awschalom, Room temperature coherent control of defect spin qubits in silicon carbide. *Nature* **479,** 84-87 (2011).

[35] M. Widmann, S-Y. Lee, T. Rendler, N.T. Son, H. Fedder, S. Paik, L-P. Yang, N. Zhao, S. Yang, I. Booker, A. Denisenko, M. Jamali, S. Ali Momenzadeh, I. Gerhardt, T. Ohshima, A. Gali, E. Janzén, J. Wrachtrup, Coherent control of single spins in silicon carbide at room temperature. *Nat. Mater.* **14**, 164-168 (2015).

[36] R. Kolesov, K. Xia, R. Reuter, R. Stöhr, A. Zappe, J. Meijer, P.R. Hemmer, J. Wrachtrup, Optical detection of a single rare-earth ion in a crystal. *Nature Commun.* **3**, 1029 (2012).

**Acknowledgements:** We thank D. Pagliero for assistance with some of the experiments. **Funding:** Support for this work was provided by the National Science Foundation through grant NSF-1314205. **Author contributions**: S.D., J.H. and H.J. conducted the experiments. C.A.M. supervised the work and wrote the manuscript. All authors discussed the results. **Competing interests**: The authors declare that they have no competing interests. **Data and materials availability**: All data needed to evaluate the conclusions in the paper are present in the paper and/or the Supplementary Materials. Additional data related to this paper may be requested from the authors.

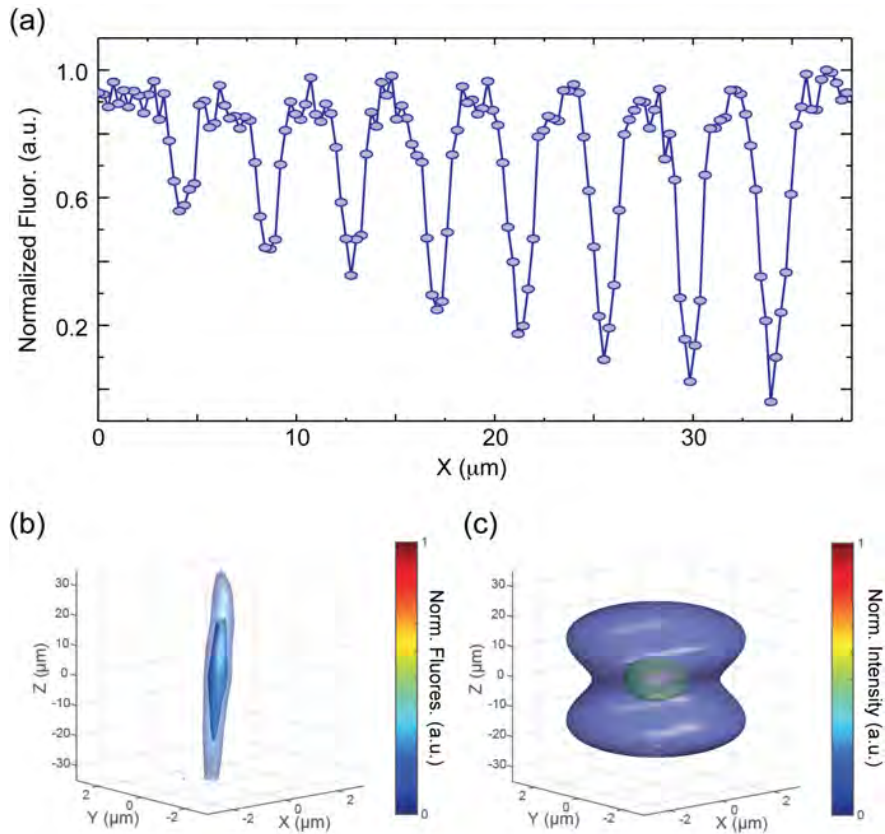

**Fig. S1. Spatial resolution of NV charge patterning.** (a) In our experiments we manipulate NV charge using a dry objective with numerical aperture *NA*=0.42. The top image shows the NV⁻ pattern resulting from parking a red beam (632 nm, 200 µW) at select locations (dark spots) separated by 4.2 µm; the exposure time increases in 10 ms steps starting from 10 ms on the left end. Readout is carried out with the same laser using a 1 ms integration time per point. A 1 mW, 532 nm laser is used initially to reset the background as explained in the main text. (b) Experimental three-dimensional isosurface plot of the ionization pattern imprinted by red laser illumination (632 nm, 200 µW, 50 ms); here the *Z*-axis coincides with the direction of beam propagation. (c) Simulated beam intensity profile for a 632 nm Gaussian laser beam focused using a 0.42 NA objective. Comparison with (b) exposes the non-linearity of the ionization process. During imaging we use a 632 nm laser for readout with a power of 200 µW and an integration time of 1 ms per pixel.

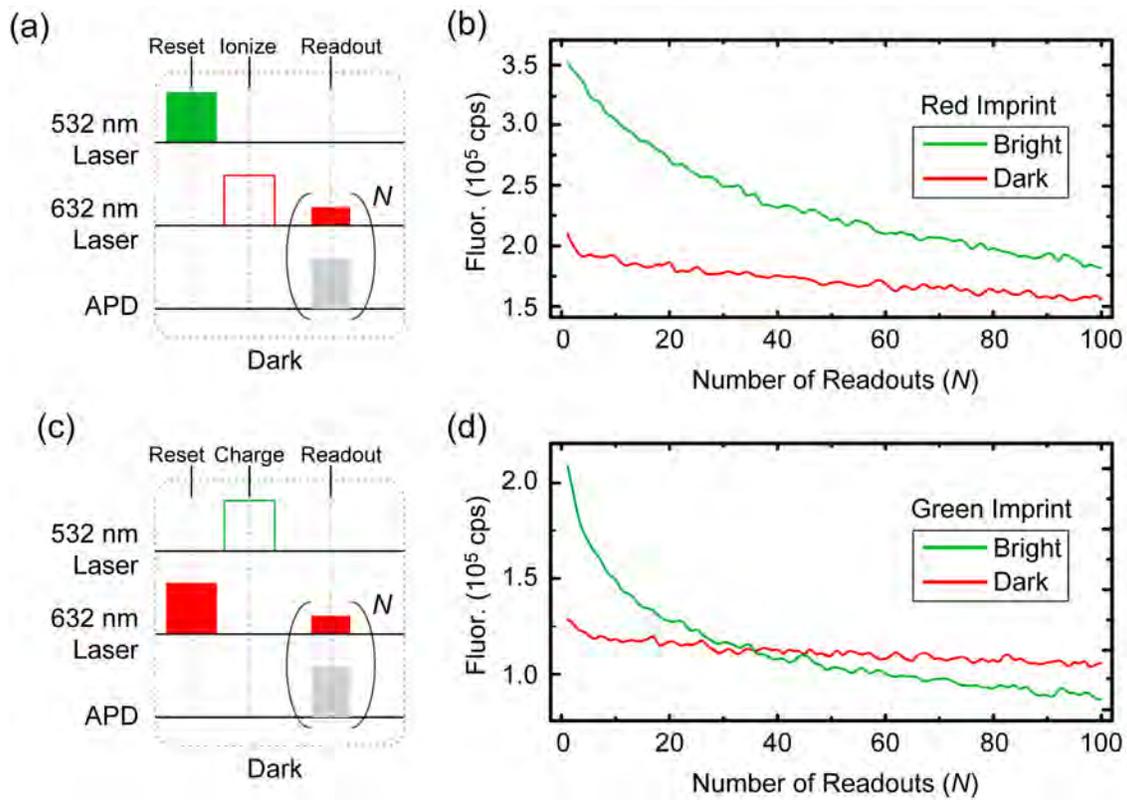

**Fig. S2. Impact of multiple readouts on NV fluorescence contrast.** (a) 'Red imprint' protocol. After a reset laser pulse (532 nm, 1 mW, 1 ms), initialization into the 'dark state' (majority of $NV^0$) is attained via red illumination (632 nm, 200 μW, 100 ms); alternatively, initialization into the 'bright state' (majority of $NV^-$) omits the red pulse. (b) NV fluorescence in the bright and dark states (green and red traces, respectively) as a function of readout cycles. (c) 'Green imprint' protocol. In this case, the bright state is produced via a red reset pulse (632 nm, 200 μW, 100 ms) followed by a green pulse (532 nm, 30 μW, 5 ms); to produce a dark initial state, the green laser pulse is skipped. (d) Same as in (b) but for a green imprint. Note the accelerated fluorescence decay as a function of the number of readouts. In (b) and (d) the readout pulse (632 nm) has a power of 200 μW and 150 μW, respectively with a duration of 1 ms.

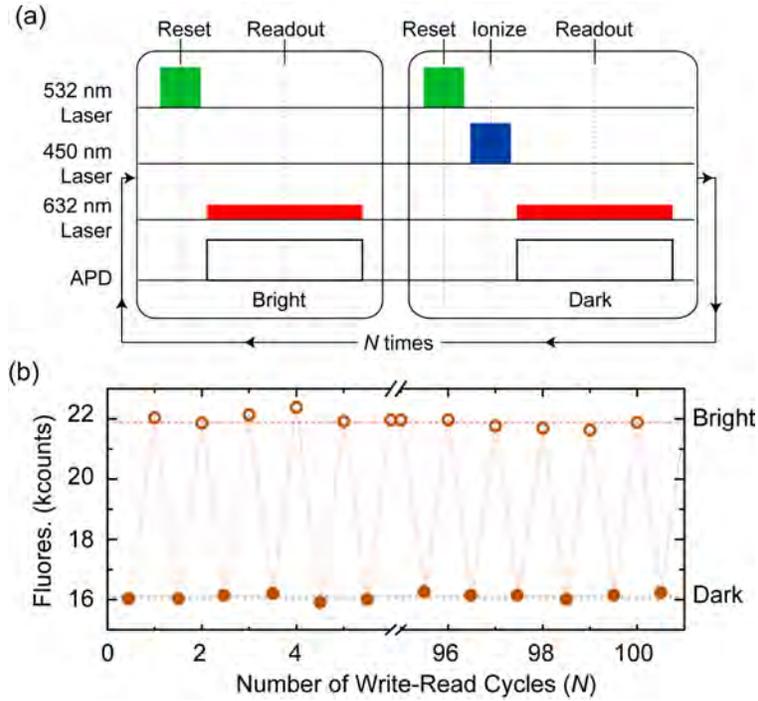

**Fig. S3. NV response upon multiple read/write cycles.** (a) Schematics of the pulse sequence. We use a green laser pulse (532 nm, 1 mW, 50 µs) or a train of femtosecond pulses (each lasting 120 fs with a repetition rate of 78 MHz at 450 nm) to locally bring the NV ensemble into the neutral or negatively charged states, respectively. The duration and average power of the femtosecond train is 30 µs and 400 µW, respectively. To probe the resulting NV charge state we collect the fluorescence created by red excitation (632 nm, 50 µW) throughout a 200 µs time interval. (b) Fluorescence after initializing into $NV^-$ (solid circles) or $NV^0$ (open circles) as a function of the number of repetition cycles $N$. We observe no change in the amplitude of the NV fluorescence contrast upon multiple ionization and recharge cycles. The solid line serves as a guide to the eye.

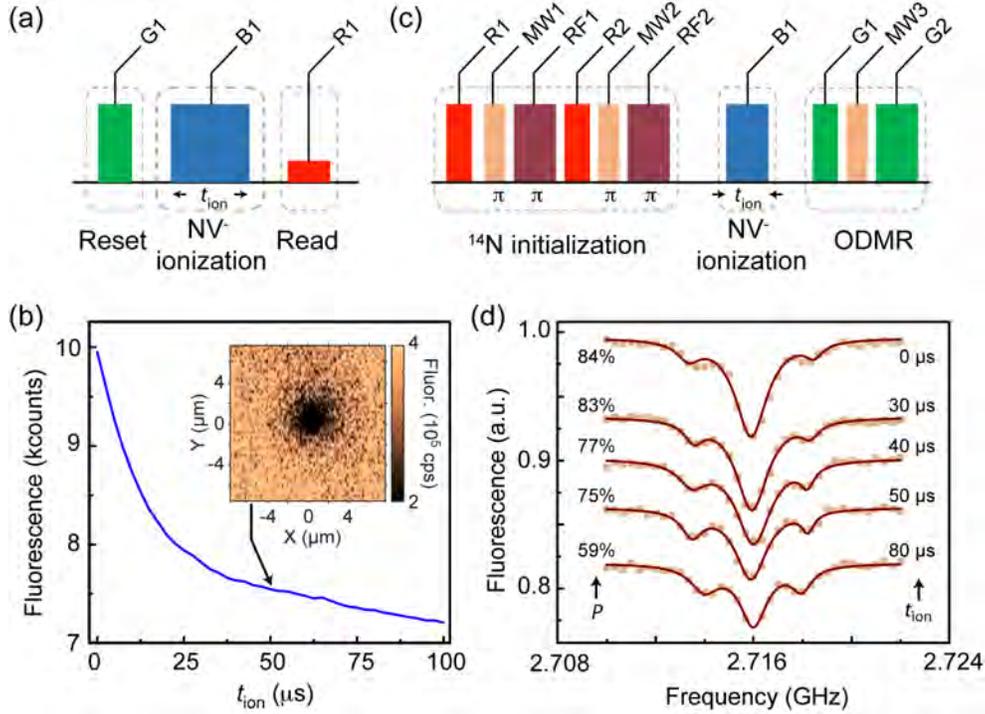

**Fig. S4. Impact of NV⁻ ionization on the ¹⁴N nuclear spin polarization.** (a) To probe the rate of NV⁻ excitation under blue excitation we readout the fluorescence from a red laser pulse R1 (632 nm, 100 µW, 100 µs) after illumination B1 at 450 nm for a variable time $t_{ion}$. After each observation, the green laser pulse G1 (532 nm, 1 mW, 100 µs) resets the NV system to the negatively charged state. Blue excitation is generated via an ultrafast laser producing 120 fs pulses at a repetition rate of 76 MHz and with average power of 400 µW. (b) NV⁻ fluorescence as a function of the ionization time $t_{ion}$. The insert shows an image around the point of blue illumination (coincident with the image center) after a ionization time of 50 µs. The image size is 100 ×100 pixels, the integration time is 1 ms per pixel and red laser power during the scan is 200 µW. (c) To assess the effect of (femtosecond) blue excitation on the nuclear spin polarization we initialize the ¹⁴N into $|m_I = 0\rangle$ and carry out pulsed ODMR spectroscopy preceded by illumination (450 nm, 400 µW) for a variable time $t_{ion}$. Recharge into NV⁻ takes place during G1 (532 nm, 1 mW, 3 µs). Photon detection is carried out during the first 500 ns of G2 (532 nm, 1 mW, 30 µs). (d) NV⁻ pulsed ODMR spectra for different ionization times; brown squares represent data points and solid lines indicate numerical fits to a background level and three Gaussians centered around the hyperfine shifted NV⁻ spin transitions. The ¹⁴N nuclear spin polarization $P$ is calculated as the ratio between the area under the central peak (corresponding to $m_I=0$) and the overall dip area. Comparison of the relative amplitudes in each hyperfine split spectrum shows virtually no change when $t_{ion}$ is sufficiently short.

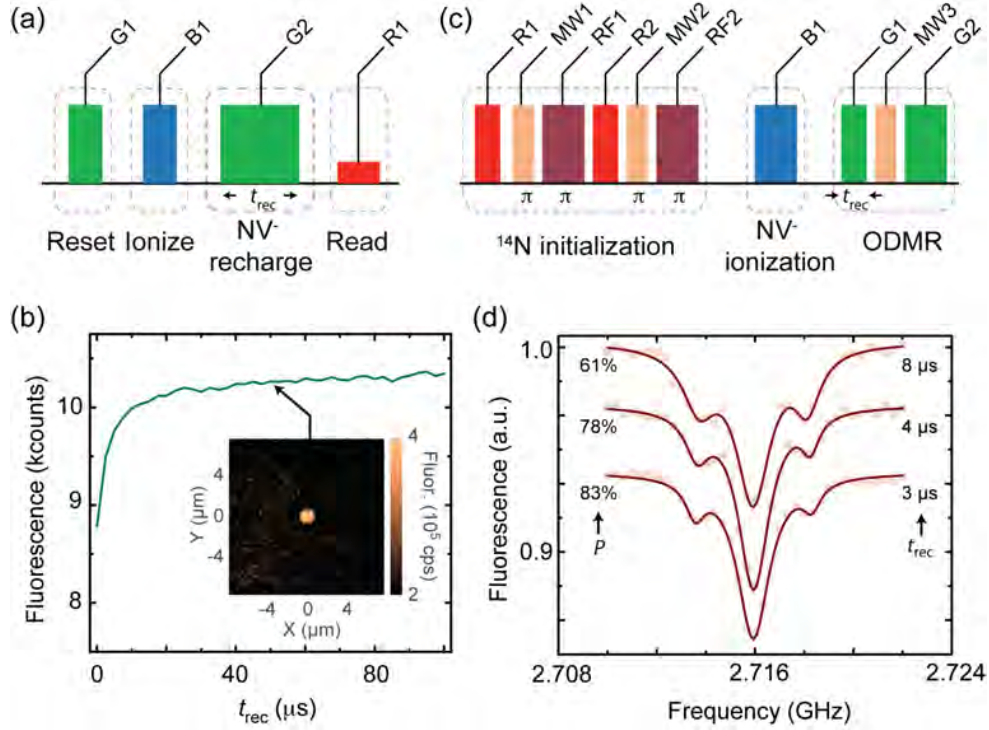

**Fig. S5. Impact of NV⁻ recharge on the ¹⁴N nuclear spin polarization.** (a) After NV⁻ depletion via a blue laser pulse (B1, 450 nm, 400 µW, 30 µs), green illumination (G2, 532 nm, 1 mW) for a variable duration $t_{rec}$ brings back NVs to the (mostly) negatively charged state. Readout is carried out with a red pulse (R1, 632 nm, 100 µW, 100 µs); the reset pulse (G1, 532 nm, 1 mW, 100 µs) ensures a well-defined initial state. (b) NV⁻ fluorescence as a function of the recharge pulse duration $t_{rec}$ upon application of the pulse sequence in (a). (c) Following initialization of the ¹⁴N spin into $m_I=0$, NV⁻ centers undergo a cycle of one-photon ionization (B1, 450 nm, 400 µW, 30 µs) and recharge (G1, 532 nm, 1 mW). Photon detection is carried out during the first 500 ns of G2 (532 nm, 1 mW, 30 µs). (d) NV⁻ ODMR is used to compare the ¹⁴N polarization at various recharge times $t_{rec}$.

## Section S1. Charge-conditioned polarization of the $^{14}$N spin

Polarization of the NV⁻ nuclear spin is carried out using a 'population trapping' scheme of the form R1-MW1-RF1-R2-MW2-RF2, where R1, R2 are red (632 nm) laser pulses, and MW1, MW2 (RF1, RF2) are microwave (radio-frequency) inversion pulses (Figs. S4 and S5). Briefly, red laser pulses initialize the NV⁻ electronic spin $S$=1 into the state $|m_S = 0\rangle$; since the ionization rate at this wavelength is low, the NV⁻ charge state remains unchanged so long as the pulse duration is sufficiently short. In this limit, polarization of the $^{14}$N nuclear spin $I$=1 into $|m_I = 0\rangle$ is the result of two successive CNOT gates each comprising two selective π-pulses: More specifically, if the NV-$^{14}$N spin system is assumed to be in the $|m_S = 0, m_I = 1\rangle$ state, then MW1 — acting selectively on the $|m_S = 0, m_I = 1\rangle \leftrightarrow |m_S = -1, m_I = 1\rangle$ transition — and RF1 — resonant with the $|m_S = -1, m_I = 1\rangle \leftrightarrow |m_S = -1, m_I = 0\rangle$ transition — produce the state $|m_S = -1, m_I = 0\rangle$, which then converts into $|m_S = 0, m_I = 0\rangle$ upon application of R2. On the other hand, starting from $|m_S = 0, m_I = -1\rangle$, MW2 — selective on the $|m_S = 0, m_I = -1\rangle \leftrightarrow |m_S = -1, m_I = -1\rangle$ transition — and RF2 — resonant with the $|m_S = -1, m_I = -1\rangle \leftrightarrow |m_S = -1, m_I = 0\rangle$ transition — drive the NV-$^{14}$N spin state into $|m_S = -1, m_I = 0\rangle$, which then transforms into $|m_S = 0, m_I = 0\rangle$ upon optical pumping with red (or green) laser light. Note that throughout the protocol all pulses (including laser pulses) have no effect on neutral NVs (featuring different optical and magnetic resonance transition frequencies), hence making the $^{14}$N spin polarization conditional on the NV charge state, i.e., the equivalent of a charge-to-spin conversion.

## Section S2. Impact of NV⁻ ionization and recharge on $^{14}$N spin polarization

To assess the influence of charge manipulation on nuclear spins, we subject spin-polarized $^{14}$N spins in negatively charged NVs to a cycle of ionization and recharge via the consecutive application of blue and green laser pulses, each having a variable duration $t_{ion}$ and $t_{rec}$, respectively (see Figs. S4 and S5). The blue laser pulse — originating from an ultrafast laser — is itself a train of 450 nm femtosecond pulses (see above). Unlike the two-photon processes presented in Fig. 1a of the main text, illumination at this wavelength ionizes NV⁻ via a one-photon absorption process. This form of ionization propels the excess electron directly into the conduction band and hence protects the $^{14}$N spin from relaxation via level mixing in the NV⁻ excited states.

During nuclear spin initialization (Figs. S4c and S5c) microwave pulses MW1 and MW2 act selectively on the transitions $|m_s = 0, m_I = +1\rangle \leftrightarrow |m_s = -1, m_I = +1\rangle$ and $|m_s = 0, m_I = -1\rangle \leftrightarrow |m_s = -1, m_I = -1\rangle$, respectively, whereas radio-frequency pulses RF1 and RF2 are tuned to the transitions $|m_s = -1, m_I = +1\rangle \leftrightarrow |m_s = -1, m_I = 0\rangle$ and $|m_s = -1, m_I = -1\rangle \leftrightarrow |m_s = -1, m_I = 0\rangle$, respectively; the red (R), and green (G) laser pulse powers (durations) are 250 µW (15 µs) for pulses R1, R2, 1 mW (3 µs) for pulse G1, and 1 mW (30 µs) during pulse G2. The average blue laser power during B1 is 400 µW.